\documentstyle[preprint,aps]{revtex}
\begin{document}
\draft
\tighten
\title{Resonant nuclear transition in the  $d\mu{^6\!Li}$ muonic molecule}
\author{
   S. A. Rakityansky \thanks{Permanent address:
    Joint Institute  for Nuclear Research, Dubna, 141980, Russia}, S.A.
     Sofianos}
\address{ Physics Department, University of South Africa,  P.O.Box
     392, Pretoria 0001, South Africa}
   \author{
    V. B. Belyaev, V. I. Korobov}
\address{ Joint Institute for Nuclear Research, Dubna, 141980, Russia}
\date{\today}
\maketitle
%%%%%%%%%%%%%%%%%%%%%%%%%%%%   ABSTRACT   %%%%%%%%%%%%%%%%%%%%%%%%%%%%%%%
\begin{abstract}
The reaction rate of the nuclear fusion
$d{\,}^6\!Li\rightarrow{\,}^8\!Be^*(2^{+},0)$
 is estimated in the case where the nuclei are confined to each other by
 a muon. For the description of nuclear transitions,
a method which is analogous to the  Linear Combination
of Atomic Orbitals has been used. Using the complex
coordinate rotation method, we found that a molecular
$d-\mu-{^6\!Li}$ state exists with energy $(-20.3084-i0.0066)$ eV and
$|\Psi_{m}(0)|=0.44\;10^{-7}\;{\rm fm}^{-\frac{3}{2}}$. The nuclear
wave functions needed, were constructed in the form of antisymmetrized
products of harmonic-oscillator functions for the three-cluster
approximation, $(dd\alpha)$, to the five-body $(NNNN\alpha)$ problem.
It was found that the reaction rate $\lambda$ is strongly dependent on the
energy gap between the $d{\,}^6\!Li$ threshold and the energy of the final
${\,}^8\!Be^*$ resonant state. The value of $\lambda$ obtained by  averaging
over the width of this resonance, is $0.183\;10^{10} {\rm sec}^{-1}$.\\

 \end{abstract}
 %%%%%%%%%%%%%%%%%%%%%%%%%%%%%%%%%%%%%%%%%%%%%%%%%%
 \section{Introduction}
%%%%%%%%%%%%%%%%%%%%%%
Low-energy interactions among atomic nuclei are of great interest from
pure theoretical as well as from  practical point of view. From the former,
they provide valuable information  on the nucleon--nucleon
interaction and on fine details concerning nuclear structure.
By this we mean the manifestation of the main symmetries of  strong
interactions, such as charge symmetry, time invariance, etc at very low
energies. Another issue here is the importance of the small components
of nuclear wave functions. Very often these small components play a crucial
role at super low energies\cite{dd}, we are interested in.
{}From the practical point of view, we mention here the importance of
 the low-energy nucleus-nucleus data in nuclear astrophysics.\\

 Nucleus-nucleus collision data are not  available at energies
  below  the keV region\cite{low}. Meanwhile nuclear astrophysics deals
 with nuclear reactions occuring inside the stellar objects at these low
energies and thus the available scattering data are usually  extrapolated
there. However, in certain systems of light nuclei, we may study
nuclear reactions at very low energies, namely, reactions
between nuclei confined inside muonic molecules. Being in a molecular state,
nuclei have rather small relative kinetic energy and any possible reaction
can be treated as a threshold reaction.\\

Nuclei in a molecule are relatively far from each other and the
probability of their collision is very small. However, in some cases this
 is partly compensated by the large value of the nucleus-nucleus
reaction cross-section which is due  to the existence of a threshold
 resonance.In this paper we consider one such case, namely,
the fusion of $d$ and $^6\!Li$ nuclei, which forms the threshold resonant
 state $^8\!Be^*(2^{+},0)$.\\

Previously, we investigated three such cases, namely the
$d\mu{^7\!Be}$, $p\mu{^{10}\!Be}$, and $t\mu{^3\!He}^*$  \cite{bfs,my,tt}.
It was found that the
reaction rates in all cases studied were rather high indicating that if such
molecules are formed, the mixing of states could generate nuclear
reactions at appreciable rate.\\

The paper is organized as follows:
In Sec. II we apply an LCAO motivated approach to derive formal
expressions for the reaction rate;
in Sec. III we briefly describe how the three-body Coulomb problem
 $d-\mu-{^6\!Li}$ is treated; Sec. IV contains details concerning
 the employed nuclear wave functions. The  potentials used and the
procedure of calculating their matrix elements are described
 in Sec. V. Finally, our results are presented and discussed in Sec. VI.

%%%%%%%%%%%%%%%%%%%%%%%%%%%%%%%%%%%%%%%%%%%%%%%%%%%%%%%%%%%%%%%%%%%%%%%%
%%%%%%%%%%%%%%%%%%%   SECTION I   %%%%%%%%%%%%%%%%%%%%%%%%%%%%%%%%%%%%%%
%%%%%%%%%%%%%%%%%%%%%%%%%%%%%%%%%%%%%%%%%%%%%%%%%%%%%%%%%%%%%%%%%%%%%%%%
\section{Reaction rate}
Consider the molecular system $d-\mu-{^6\!Li}$  in its $S$-wave ground
state. Using the complex coordinate rotation  method (see forth
Sec. III), we found  that a quasistationary state exists with energy
 $(-20.3084 -i 0.0066)$ eV which is in accordance with the
early estimates of -19.8 eV \cite{krav} and -23.8 eV \cite{bel}.\\

This binding energy is negligible on the nuclear energy scale, and
beyond the typical accuracy with which the nuclear spectra are known.
Thus, we can assume that the nuclear subsystem,  $d\,{^6\!Li}$,  of the
above molecule, is close to its threshold energy.
This threshold lies  22.2798 MeV above the ${^8\!Be}$
ground state i.e. almost at the same energy where the ${^8\!Be}$ nucleus
 has an excited state at $(22.2-i 0.8){\rm MeV}$ \cite{spectr} with
 quantum numbers $(J^\pi,T)=(2^{+},0)$. This state  can be
formed also from the corresponding nuclear quantum numbers
 $d(1^{+},0), \,{^6\!Li}(1^{+},0)$, and zero nucleus-nucleus
angular momentum. Therefore, the nuclear subsystem $d\,{^6\!Li}$
 of the molecule $d-\mu-{^6\!Li}$ can be fused via the reaction
\begin{equation}
\label{reac}
       d+{^6\!Li}\longrightarrow{^8\!Be}^* +79.78\, {\rm keV}.
\end{equation}
To calculate the reaction rate of this process we employ a method which
is motivated by the so-called Linear Combination
of Atomic Orbitals (LCAO) \cite{lcao} approach. We have used this method
previously to study the
      $d-\mu-{^7\!Be}$ \cite{bfs} and $p-\mu-{^{10}\!Be}$ \cite{my}
systems.\\

Following the LCAO approach, we write the total Hamiltonian $H$,
of the nine-body system considered here, in two equivalent forms
corresponding to the initial and final clustering
\begin{eqnarray}
    H=H_1+V^s,\label{Hh1}\\
    H=H_2+H^{\mu},\label{Hh2}
\end{eqnarray}
where
\begin{equation}
     H_1 = H_d+H_{^6\!Li}+H_0+V^c+H^{\mu},\label{Hch1}
\end{equation}
and
\begin{equation}
     H_2 = H_{^8\!Be}.\label{Hch2}
\end{equation}
are the `unperturbed' or `channel' Hamiltonians.
In these expressions, $H_d$, $H_{^6\!Li}$,
and $H_{^8\!Be} $ are the nuclear hamiltonians of the clusters;
 $H_0$ is the kinetic energy operator describing the
nucleus-nucleus relative motion while
$V^s$ is part of the strong potentials not included in $H_d$ and
 $H_{^6\!Li}$, i.e it represents interactions among  nucleons belonging to
different clusters; $V^c$ is the nucleus-nucleus Coulomb
potential and $H^{\mu}$  consists of the muon kinetic energy operator
and its Coulomb interaction with the nuclear particles.\\

The `unperturbed' Hamiltonians $H_1$ and $H_2$ have eigenstates
     $\Psi_1={\cal A}\lbrace\Psi_d\Psi_{^6\!Li}\Psi_{mol}\rbrace$
 and
     $\Psi_2=\Psi_{^8\!Be^*}\phi_\mu$ ,
satisfying
\begin{equation}
   \label{HPsii}
    H_i\Psi_i=E_i\Psi_i,\qquad i=1,2\,
\end{equation}
and eigenenergies $E_1$ and $E_2$  which are very close to each other.
In the above $\Psi_{mol}$ is the  three-body wave function of the molecular
state; $\phi_\mu$ is the atomic wave function of the muon in the
Coulomb field  of ${^8\! B}e$ and  $\Psi_d,\,\Psi_{^6\!Li}$,
 and $\Psi_{^8\!Be^*}$ are the fully antisymmetric wave functions of the
 nuclear subsystems while ${\cal A}$ is the antisymmetrizer involving
nucleons from different nuclei.\\

In this system, the muon merely acts as a mediator creating an effective
 force which holds the nuclear clusters, $d$ and $^6\!Li$, together at
molecular distances. In order to gain a qualitative insight into the problem,
one can assume that the effective interaction between $d$ and $^6\!Li$ is
as schematically shown  in Fig. 1. When the clusters are close
to each other they merge into $^8\!Be^*$ which can be considered as
 an eigenstate of the deep well. When they are separated at large
 distances, they can occupy the molecular  eigenstate of the shallow
 well. The basic feature of this system is that the eigenvalues  of these
 two eigenstates $E_1$ and $E_2$ are practically the same. As a result,
the molecular and the compound nucleus states can be mixed. Despite the
simplicity of such a two-cluster picture, we expect that the mixing  of the
two states can in reality take place  as well.\\

The `perturbations' $V^s$ and $H^\mu$ mix the `unperturbed' solutions
$\Psi_1$ and $\Psi_2$ and shift the eigenenergies $E_1$ and $E_2$. Thus,
from the above arguments and following the LCAO approach, we look for
a solution of the Schr\"odinger equation
\begin{equation}
\label{HPsi}
           H\Psi=E\Psi
\end{equation}
in the form of a linear combination
\begin{equation}
\label{PCC}
        \Psi=C_1\Psi_1+C_2\Psi_2.
\end{equation}
Using Eq. (\ref{PCC}) in Eq. (\ref{HPsi}) and projecting with
$\langle \Psi_i |$, we obtain the  following   algebraic
linear equations for the coefficients  $C_i$ \cite{my}
\begin{eqnarray}
        C_1(H_{11}-E)&+C_2(H_{12}-EI)&=0,\nonumber\\
          \phantom{-}\label{CIS}\\
        C_1(H_{21}-EI^*)&+C_2(H_{22}-E)&=0,\nonumber
\end{eqnarray}
where $H_{ij}\equiv\langle\Psi_i|H|\Psi_j\rangle$ and
$I=\langle\Psi_1|\Psi_2\rangle$, the overlapping integral.
The determinant of the degenerated system (\ref{CIS}) is equal to
zero for the two values of the energy-parameter
\begin{eqnarray}
          E^{(\pm)}={\displaystyle \frac{1}{2(1-|I|^2)}}
           \biggl \{ H_{11}+H_{22}-
           (IH_{21}+I^*H_{12})\pm \biggr.
 &&\nonumber\\
\phantom{-}\label{EPM}\\
            \biggl. \sqrt{\lbrack
            H_{11}+H_{22}-(IH_{21}+I^{\ast}H_{12})\rbrack^2-4(1-|I|^2)
            (H_{11}H_{22}-H_{12}H_{21})}\, \biggr\}&&,\nonumber
 \end{eqnarray}
 which are the `perturbed' energy levels corresponding to the
 `unperturbed' values $E_1$ and $E_2$.\\

To find out which of the two levels $E^{(+)}$ and $E^{(-)}$  stems
from the molecular and which from the nuclear state, we modify
Eq. (\ref{EPM}) by  writing
\begin{eqnarray}
        H_{11}&=&E_1\phantom{I^*} +V_{11}^s,\label{15}\\
        H_{12}&=&E_1I\;\, +V_{12}^s,\label{16}\\
        H_{21}&=&E_1I^* +V_{21}^s,\label{17}\\
        H_{22}&=&E_2\phantom{I^*} +H_{22}^\mu,\label{18}
\end{eqnarray}
where $V_{ij}^s=\langle\Psi_i|V^s|\Psi_j\rangle$ and
  $H_{22}^\mu=\langle\Psi_2|H^\mu|\Psi_2\rangle$.
Noting that $(1-|I|^2)^{-1}=1+g$, where $g=|I|^2+|I|^4+|I|^6+\dots$ is a
small quantity, we get
\begin{eqnarray}
\nonumber
     E^{(+)} &=& E_1+\Delta_1,\\
\phantom{-}\label{EPMD}\\
     E^{(-)}&=&E_2+\Delta_2,
\nonumber
\end{eqnarray}
where
\begin{eqnarray}
\nonumber
         \Delta_1&=&V_{11}^s+gH_{11}+(1+g)\left [ (\sqrt{1+s}-1)
         (E_1-E_2+V_{11}^s-H_{22}^\mu)-R\!e(IH_{21})\right],\\
\nonumber
         \Delta_2&=&H_{22}^\mu+gH_{22}-(1+g)\left [(\sqrt{1+s}-1)
         (E_1-E_2+V_{11}^s-H_{22}^\mu)+R\!e(IH_{21})\right],
\end{eqnarray}
and
$$
    s={\displaystyle\frac{4}{(E_1-E_2+V_{11}^s-H_{22}^\mu)^2}}\left[|I|^2
    H_{11}H_{22}+|H_{21}|^2-(H_{11}+H_{22})R\!e(IH_{21})+\left(I\!m(IH_{21})
\right)^2\right].
$$
It is clear that $E^{(+)}$ is the `perturbed' molecular energy
level, and $\Delta_1$ is the shift caused by the strong interaction between
the nuclei, while $\Delta_2$ is a negligible shift of the nuclear level
caused by the muon cloud surrounding the $^8\!Be^*$ nucleus.\\

{}From the above expressions for $\Delta_1$ and $\Delta_2$ it is seen that
our approach differs from the simple perturbation theory where
$\Delta_1\approx V_{11}^s$ and $\Delta_2\approx H_{22}^\mu$.
The LCAO wave function is, therefore, written
\begin{equation}
        \Psi^{(+)}=C_1^{(+)}\Psi_1+C_2^{(+)}\Psi_2,
\label{Psi+}
\end{equation}
where $C_1^{(+)}$ and $C_2^{(+)}$ are solutions of (\ref{CIS})
corresponding to the energy  $E=E^{(+)}$.\\

The probability $P$ for the transition (\ref{reac}) is
obtained as usual by the projection
\begin{equation}
           P=\left|\langle\Psi_2|\Psi^{(+)}\rangle\right|^2.
\label{Prob}
\end{equation}
The  $P$ can be interpreted as the transition probability  through the
potential barrier. It is noted that the total wave function
           $\Psi^{(+)}\exp(-iE^{(+)}t/\hbar)$
 oscillates at the barrier with frequency $\nu=|E^{(+)}|/2\pi\hbar$.
Hence, in order to obtain the reaction rate $\lambda$, we must multiply
the probability  $P$ by $\nu$. Solving the algebraic system (\ref{CIS}),
we finally get
\begin{equation}
           \lambda={\displaystyle\frac{|E^{(+)}|}{2\pi\hbar}
                    \frac{|fI^*+1|^2}{|f|^2+fI^*+f^*I+1}},
\label{lambda}
\end{equation}
where
\begin{equation}
       f={\displaystyle\frac{E^{(+)}I-H_{12}}{H_{11}-E^{(+)}}}
\label{freq}
\end{equation}
Therefore to calculate the reaction rate we need the matrix elements
$V_{11}^s$, $V_{12}^s$, $V_{21}^s$, $H_{22}^\mu$, and the overlapping
integral $I=\langle\Psi_1|\Psi_2\rangle$. This in turn requires the
construction of the `unperturbed' molecular $(d-\mu-{^6\!Li})$
and nuclear $(^8\!Be^*)$ wave functions $\Psi_1$ and $\Psi_2$ respectively.

%%%%%%%%%%%%%%%%%%%%%%%%%%%%%%%%%%%%%%%%%%%%%%%%%%%%%%%%%%%%%%%%%%%
%%%%%%%%%%%%%%%%%   MOLECULAR STATE %%%%%%%%%%%%%%%%%%%%%%%%%%%%%%%
%%%%%%%%%%%%%%%%%%%%%%%%%%%%%%%%%%%%%%%%%%%%%%%%%%%%%%%%%%%%%%%%%%%

\section{Molecular state}
 In the initial state, the nuclei $d$ and $^6\!Li$ are very far apart
 and held by the muon at  molecular distances (in the shallow well of
Fig. 1). Since this separation is much greater than the nuclear sizes, we
consider them as point-like charged particles interacting via a pure
Coulomb potential. In order to solve the resulting three-body
$(d\mu{^6\!Li})$ Coulomb problem, we employ the complex coordinate rotation
method which has been successfully used in atomic physics for calculating
positions and widths of resonant states ( see, for example,
Ref.\cite{Ho} and references therein ).\\

The main idea of this method consists in the complex rotation of the spatial
coordinates
\begin{equation}
\label{crot}
\vec r\rightarrow\vec r\exp(\theta),\qquad\theta=\theta_r+i\theta_i,
\end{equation}
which makes the resonance state wave function square integrable. This,
 in turn, permits the use of variational methods for their location since
the rotation leaves the resonance pole positions intact. Another advantage
of this approach is that the search for resonances as eigenvalues in the
complex plane has a rigorous mathematical foundation\cite{BalCom,Simon}.\\

Muonic molecular systems have extremely narrow resonances, with  widths
which are  difficult to calculate, because the accuracy of the
numerical calculation for the complex energy  has to be at least
higher than the width magnitude. A variational method based on  a
random-tempered Slater-type exponential expansion exhibited high
convergency and accuracy in calculations of muonic molecular bound states
\cite{FroEfr,AleMon}. Here we demonstrate that this
 method is also successful in the case of $d\mu{^6\!Li}$ system.\\

The system of interest consists of three particles, a negative muon of mass
$m_{\mu}$ and two nuclei of masses $M_a$ and $M_b$, where $a$, hereafter,
stands for a lithium nucleus and $b$ for the deuteron.
The Hamiltonian (in muonic atomic units $e=\hbar=m_{\mu}=1$), after
separating the center of mass motion, can be written as
\begin{equation}\label{Hamiltonian}
H_{mol} = -{1\over2m_a}\Delta_{{\vec r}_a}-{1\over2m_b}\Delta_{{\vec r}_b}
-\nabla_{{\vec r}_a}\cdot \nabla_{{\vec r}_b}
-{3\over r_a}-{1\over r_b}+{3\over R}\equiv T+V,
\end{equation}
where ${\vec r}_a$ and ${\vec r}_b$ are the vectors towards the muon from the
two nuclei, $R$ denotes the distance between the nuclei, and
$m_i=m_{\mu}M_i/(m_{\mu}+M_i)$
are the reduced masses of the respective muonic atoms ($i=a,b$).
This set of coordinates, $\vec r_a,\vec r_b$, and $\vec R$, is more
appropriate to handle the molecular three-body problem than the Jacobi
coordinates $(\vec R,\vec r_\mu)$ commonly used in few-body problems.\\

In the complex coordinate rotation theory the resonant state is defined as
the solution of the eigenvalue problem
\begin{equation}\label{eigen}
(H_{mol}(\theta)-E)\Psi_{mol}=0,\qquad H_{mol}(\theta) = U(\theta)H_{mol}
U^{-1}(\theta),
\end{equation}
for the Hamiltonian $H_{mol}(\theta)$ analytically continued into the
complex plane of dilation parameter $\theta$ of the transformation
(\ref{crot}).
The corresponding transformation  of the  molecular states in the
Hilbert space  is defined via

\begin{equation}\label{dilatation}
\left[U(\theta)F\right]({\vec r}) = e^{3\theta/2}F(e^{\theta}{\vec r}).
\end{equation}
Such a transformation has a great computational advantage for systems with
Coulomb interactions. The kinetic and potential parts scale as
$\exp{(-2\theta)}$
and $\exp{(-\theta)}$, respectively, and the Hamiltonian can be written as
\begin{equation}\label{DHamiltonian}
H_{mol} = Te^{-2\theta}+Ve^{-\theta}.
\end{equation}

The variational wave function for the state with total orbital angular
momentum $L=0$ is expanded in the form
\begin{equation}\label{expansion}
\Psi_{mol}({\vec R},{\vec r}_\mu) =
\sum_{i=1}^{N} c_i\exp(-\alpha_ir_a-\beta_ir_b-\gamma_iR),
\end{equation}
where the vector $\vec r_\mu$ directed from the center of mass of the
nuclear subsystem to the muon, is the linear combination
$$
       \vec r_\mu={\displaystyle \frac{M_a\vec r_a + M_b\vec r_b}
      {M_a+M_b} }
$$
of $\vec r_a$ and $\vec r_b$.
The nonlinear parameters are chosen by a pseudorandom algorithm, previously
 used by Thakkar and Smith \cite{ThaSmi}, and are generated by the formulae
\begin{equation}\label{tempering}
\begin{array}{r@{}l}
\alpha_i&=\hbox{A}\langle\sqrt{2}[i(i+1)/2]\rangle,\\
\beta_i&=\hbox{B}\langle\sqrt{3}[i(i+1)/2]\rangle, \\
\gamma_i&=\hbox{C}\langle\sqrt{5}[i(i+1)/2]\rangle,
\end{array}
\end{equation}
where $\langle\dots\rangle$ denotes the fractional part of a number.
The tempering parameters A, B, and C can be taken more or less arbitrarily.
A resonable requirement is that these parameters must be chosen to
reflect the geometry of the molecule.\\

Once the nonlinear parameters $\alpha_i$, $\beta_i$, and $\gamma_i$
are fixed, the coefficients $c_i$  can be found by
solving the complex nonhermitian matrix eigenvalue problem
\begin{equation}\label{gen.eigen}
{\bf A}{\bf c}_k = E_k{\bf B}{\bf c}_k,
\end{equation}
where
$$
{\bf A} = \langle\Psi_{mol}|H_{mol}(\theta)|\Psi_{mol}\rangle,\quad
{\bf B} = \langle\Psi_{mol}|\Psi_{mol}\rangle.
$$
The corresponding discrete complex eigenvalues have the form
\begin{equation}\label{c.eigen}
E_k = E_r-i\Gamma_m/2
\end{equation}
where $E_r$ gives the position and $\Gamma_m$  the width of the resonance. \\

It should be emphasized that the exact eigenvalues do not depend on the
rotating parameter $\theta$. However, employing the ansatz (\ref{expansion}),
we deal with approximate eigenvalues which, therefore, have such dependence.
On the other hand, in the area near the spectral points
(including resonances) the dependence of eigenvalues on any parameter
 should  be weak since the corresponding functional must be stable.
 Since  the variational parameters of the expansion (\ref{expansion}) are
fixed by (\ref{tempering}) and  (\ref{gen.eigen}), then the only
parameters which remain free  are the real and imaginary parts of
dilation angle $\theta$. Therefore, the resonance positions and widths are
deduced from the condition that a discrete complex eigenvalue is stabilized,
i.e
\begin{equation}
        {\partial E_k\over\partial\theta}=0,
\end{equation}
with respect to variations of  the complex dilation parameter $\theta$.\\

To solve Eq.~(\ref{gen.eigen}) numerically, the inverse
iteration method adapted to symmetric complex matrices has been employed.
This method is very effective, as far as computational effort is
concerned, and  stable to round-off errors.
This enables us to extend the number of the expansion functions up
to 1400 terms.\\

In our approach,  the full three-body wave function  $\Psi_{mol}$ is
not needed since in  actual calculations of the fusion rate, only the
effective two-body wave function  $\Psi_m(\vec R)$ describing the
nucleus-nucleus relative motion is required. Moreover,
$\Psi_m$ will appear only in the integral expressions involving either  the
short-range strong potentials or the nuclear wave-functions localized within
a volume extending to few fm. Hence, such integrals take into account  only
a small area around the  $R=0$ point. The boundary value $\Psi_m(0)$ is
obtained from $\Psi_{mol}$ by noting that after the complex rotation,
the resonance wave function is square integrable and thus, the density
 $\rho_N^{(\theta)}(0)$ of the probability for the nuclei to be at
vanishing internuclear separation, $R=0$, is given by
\begin{equation}\label{dzero}
\rho_N^{(\theta)}(0) = {\displaystyle\frac{\int\,d\vec r_\mu\,\Psi_{mol}^
{\dagger}(0,\vec r_\mu) \Psi_{mol}(0,\vec r_\mu)}
{\langle\Psi_{mol}|\Psi_{mol}\rangle}}\;.
\end{equation}
where $\Psi_{mol}^{\dagger}$ is the solution of the adjoint equation
\begin{equation}
\Psi_{mol}^{\dagger}\,H_{mol}(\theta) = \Psi_{mol}^{\dagger}E^*.
\end{equation}
%%%%%%%%%%%%%
Finally, the rotated value of $\rho_N^{(\theta)}(0)$ can be easily
transformed to $\theta=0$ via
\begin{equation}\label{extrapolation}
\rho_N^{(0)}(0)=e^{-3\theta}\rho^{(\theta)}_N(0).
\end{equation}
Therefore, the boundary value $\Psi_m(0)$  is
$$
|\Psi_m(0)|=\sqrt{\rho_N^{(0)}(0)}.
$$
Our calculations were performed in a 32 digits arithmetics.
The values of $M_{{}^6\!{Li}\,}$, $M_d$ and $m_{\mu}$
are (in units of electron mass) 10961.9, 3670.481 and 206.7686,
respectively, while the values of the tempering parameters  used are
 A=1.8, B=1.8, and C=2.8.

In Fig. 2 we present the results obtained for an eigenvalue as a
 function of the complex dilation parameter and with variational wave
function  (\ref{expansion}) having a  total number of terms $N=900$.
The curves correspond to different values of the parameter
$\alpha=e^{\theta_r}$. Each curve connects points on the
rotational path for the fixed value of $\alpha$ and different values of
 the rotational angle $\theta_i=0,0.02,\dots,0.2$ in  steps of
 $\Delta\theta_i=0.02$.  It is clearly seen that the rotational paths
form the stabilized eigenvalue of $(-20.3084-i0.0066)$eV at
 $\theta_i\approx0.1$, the accuracy being of the order of $10^{-4}$eV.
The $\Gamma_m$  corresponds to the predissociation rate of
$1.8\!10^{13}\!{\rm sec}^{-1}$.
The boundary value of the nucleus-nucleus wave-function obtained by the
above procedure is
$$
|\Psi_m(0)|=0.44\;10^{-7}\;{\rm fm}^{-3/2}\,.
$$
\\
%%%%%%%%%%%%%%%%%%%%%%%%%%%%%%%%%%%%%%%%%%%%%%%%%%%%%%%%%%%%%%%%%%%%%%%%
%%%%%%%%%%%%%%%%%%%%%%%%%%%%%%%%%%%%%%%%%%%%%%%%%%%%%%%%%%%%%%%%%%%%%%%%
\section{Nuclear models}
%%%%%%%%%%%%%%%%%%%%%%%%
In order  to  reduce the eight-body nuclear system, $d\,{^6\!L}i$,
in the initial channel to manageable proportions,
$^6\!Li$ is considered to be a $(d\alpha)$-bound state system.
In this way we have in the initial state a three cluster system,
namely, the $d d \alpha$ one. It is known that such a two-body assumption
for the  $^6\!Li$ ground state, is sufficiently accurate \cite{wt}.
 This model is also consistent with the treatment of the final state of the
transition $d\,{^6\!Li}\rightarrow{^8\!Be^*}$, since when $^8\!Be$ nucleus is
 excited at $\sim20{\rm MeV}$, then only one of the two $\alpha$-clusters
in it is destroyed \cite{wt} forming two deuterons. This is
schematically depicted in Fig. 3 together with two sets of possible Jacobi
coordinates.\\

The complexity of the problem, especially the
required multidimensional integrals (up to dimension of 10), necessitated
the use  of bound state wave functions with simple radial dependence
which, however, are widely employed in the literature. These wave
 functions describe reasonably well the nuclear sizes which are crucial
 in our model. We shall discuss them below  in more details.

\subsection{The deuteron wave function}
%%%%%%%%%%%%%%%%%%%%%%%%%%%%%%%%%%%%%%
This is taken to be of Gaussian form
\begin{equation}
          \Psi_d(\vec r)={\displaystyle\frac{N_d}{\sqrt{4\pi}}\exp
          \left(-\frac{d}{2}r^2\right)\chi^s\eta^a},
\label{Psid}
\end{equation}
with
$$
         {\displaystyle d=\frac{3}{8\langle  r_d^2\rangle}},
$$
and
$$
         N_d^2=\frac{4} {\sqrt{\pi}}d^{\frac32}.
$$
where $\chi^s$ and $\eta^a$ are the spin symmetric and isospin antisymmetric
functions. The mean square radius is fixed by
$\langle r_d^2\rangle=(1.956\,{\rm fm})^2$ \cite{deutr}.\\

\subsection{The $^6\!Li$ wave function}
%%%%%%%%%%%%%%%%%%%%%%%%%%%%%%%%%%%%%%%
This is constructed from the product of $\Psi_d$
and the harmonic oscillator $2s$-wave function, describing the motion of  the
$\alpha$ particle with respect to the center of mass of the deuteron.
It can be shown
that, in the six-nucleon oscillator model of $^6\!Li$-ground state,
the degrees of freedom associated with the variable $\vec r_2$
(see Fig.3),  has two quanta of excitation  \cite{wt}. This is a
 consequence of the Pauli principle. Therefore, by using the $2s$-function
 we, indirectly, take into account effects of the Pauli principle
despite the fact that the $\alpha$-particle is treated as elementary.
Therefore the following ansatz was used
\begin{equation}
    \Psi_{^6\!Li}(\vec r_1,\vec r_2)=\Psi_d(\vec r_1)\phi_{2s}(\vec r_2),
\label{Psili}
\end{equation}
with  the $2s$--function having the usual form
\begin{equation}
    \phi_{2s}(\vec r_2)={\displaystyle\frac1{\sqrt{4\pi}}
                \frac{\sqrt{\frac83}}
      {\pi^{\frac14}a^{\frac32}}\left(\frac{r_2^2}{a^2}-\frac32\right)
      \exp\left(-\frac{r_2^2}{2a^2}\right)}\,.
\label{phi2s}
\end{equation}
The oscillator parameter $a$,
$$
      a^2={\displaystyle\frac{9}{14}}\left(\langle r_{^6\!Li}^2\rangle-
            \langle r_d^2\rangle\right),
$$
 is chosen to reproduce the experimental mean square radius of  $^6\!Li$,
 $\langle r_{^6\!Li}^2\rangle=(2.54\,{\rm fm})^2$  \cite{wt}.\\

\subsection{ The $^8\!Be^*(2^{+},0)$ wave function}
%%%%%%%%%%%%%%%%%%%%%%%%%%%%%%%%%%%%%%%%%%%%%%%%%%%
As mentioned earlier, the wave function of $^8\!Be^*(2^{+},0)$ state,
is assumed to be  that of the five-body system  $(nnpp\alpha)$.
It is well known that the ground state of $^8\!Be$ can be described
  reasonably well by the two $\alpha$--cluster model \cite{wt}.
Using the eight-body harmonic oscillator model, however, this ground state
 is a $(1s)^4(1p)^4$--configuration having four quanta of excitation.
On the other hand, in the $\alpha-\alpha$ model
all four quanta must be attributed to the $\alpha-\alpha$ relative
motion because each of the two $\alpha$-particles is in the
$(1s)^4$--configuration. Hence, for the ground state of $^8\!Be$ the
function ${\cal A}\{\Psi_{\alpha_1}\Psi_{\alpha_2}\phi_{3s}\}$ is expected
to be a good approximation. Here
$$
     \phi_{3s}(\vec r)={\displaystyle\frac1{\sqrt{4\pi}}\frac{2\sqrt{30}}
     {\pi^{\frac14}c^{\frac32}}\left[\frac14-\frac13\left(\frac{r}{c}
             \right)^2+
     \frac1{15}\left(\frac{r}{c}\right)^4\right]\exp(-\frac{r^2}{2c^2})}
$$
is the normalized harmonic-oscillator function having two  nodes and
describing the relative $\alpha-\alpha$ motion with  four quanta of
excitation. The oscillator parameter $c$ for this function,
$$
      c^2={\displaystyle\frac8{11}\left(\langle r_{^8\!Be}^2
        \rangle-\langle r_
       \alpha^2\rangle\right)},
$$
 is chosen to reproduce the experimental size
$\langle r_{Be}^2\rangle=(2.39\,{\rm fm})^2$ of Berillium
nucleus  \cite{be8}. For $\langle r_\alpha^2\rangle$ we have used the value
$(1.671 \,{\rm fm})^2$ \cite{alpha}.\\

At 22.2 MeV above the ground state, the whole excitation energy is used
 to break one of the $\alpha$--clusters \cite{wt}. It is therefore
 natural to assume that the other  $\alpha$ particle
remains  in the same $3s$-state with respect to the center of mass of the
destroyed cluster and bears the whole excitation.
Consequently the  excited  $\alpha$ cluster  should have
quantum numbers corresponding to the  $(2^+,0)$ state.
The positive parity is obtained if we have two quanta of excitation
in the oscillator model. The simplest configuration in
 this case is the  $dd$-system with relative motion wave function
 $\phi_{2s}$. The  oscillator parameter in this case is adjusted in order to
 get the correct value of the excitation energy.\\

Therefore, the wave function of the resonant  state $^8\!Be^*(2^{+},0)$ is
\begin{equation}
      \Psi_{^8\!Be^*}(\vec r_1,\vec r_4,\vec r_5,\vec r_6)={\cal N}_2
      \phi_{3s}(\vec r_6)
       {\cal A}\{\Psi_d(\vec r_1)\Psi_d(\vec r_4)\phi_{2s}(\vec r_5)
       \xi_{(12)(34)}\},
\label{PsiBe*}
\end{equation}
where
\begin{equation}
       \xi_{(12)(34)}=\eta_{12}^a\eta_{34}^a\sum_{\sigma\sigma'}
       \langle 1\sigma1\sigma'|2,\sigma+\sigma'\rangle
       \chi_{12}^s(\sigma)\chi_{34}^s(\sigma')
\label{xi1234}
\end{equation}
is the four-nucleon spin-isospin wave function constructed from the
antisymmetric isospin functions $\eta_{ij}^a$ and the symmetric spin
functions $\chi_{ij}^s$
of the $ij$-nucleon pair and ${\cal N}_2$ is the normalization constant;
 $ {\cal A}$ is the antisymmetrizer consisting of the
following permutations
\begin{equation}
\label{22}
       {\cal A}=(12)(34)-(32)(14)-(42)(31)-(13)(24)-(14)(32)+(34)(12).
\end{equation}
It is known that the antisymmetrization significantly reduces the
differences among the  various cluster representations of nuclei \cite{wt}.
 Moreover, in the harmonic oscillator model all possible cluster
representations, after being antisymmetrized, become equivalent.
This give us reasons to believe  that our wave function for $^8\!Be^*$,
constructed in the cluster representation $dd\alpha$, is a reliable one.\\

\subsection{The final and initial states}
The  wave function $\Psi_{^8\!Be^*}$ appearing in the final state
$\Psi_2$ can be obtained from (\ref{PsiBe*}).
After carrying out the normalization and antisymmetrization it is written
as follows
\begin{eqnarray}
\nonumber
    \Psi_{^8\!Be^*}(\vec r_1,\vec r_4,\vec r_5,\vec r_6)
 &=& {\displaystyle\left[3-\frac{144}{\sqrt{2}}d^{\frac32}
               b^3\frac{(1-2db^2)^2}
        {(1+2db^2)^5}\right]^{-\frac12}}\\
\nonumber
  & &\phi_{3s}(\vec r_6) \biggl \{  \Psi_d(\vec r_1)\Psi_d(\vec r_4)
      \phi_{2s}(\vec r_5)\xi_{(12)(34)}\biggr.\\
 \nonumber
  & &      -{\displaystyle\Psi_d(\frac{\vec r_1}{2}
             -\vec r_5 +\frac{\vec r_4}{2})
         \Psi_d(\frac{\vec r_1}{2}+\vec r_5 +  \frac{\vec r_4}{2})
        \phi_{2s}(\frac{\vec r_1 - \vec r_4}{2})
            \xi_{(32)(14)}}\\
\nonumber
   & &- \biggl.   {\displaystyle\Psi_d(\frac{\vec r_1}{2}
  \vec r_5 -\frac{\vec r_4}{2})
        \Psi_d(\frac{\vec r_1}{2}+\vec r_5 -\frac{\vec r_4}{2})
        \phi_{2s}(\frac{\vec r_1 + \vec r_4}{2}) \xi_{(13)(24)}}\biggr\}.
\end{eqnarray}
 The initial state wave function
         $\Psi_1={\cal A}\{\Psi_d\Psi_{^6\!Li}\Psi_{mol}\}$
can be similarly antisymmetrized with the help of the operator given
in Eq. (\ref{22}). Thus, after some algebra, we obtain
\begin{eqnarray}
      \Psi_1&(&\vec r_1,\vec r_2,\vec r_3,\vec r_4,\vec r_\mu)=
\nonumber\\
 &&{\cal N}_1\left\{ {\Psi_d(\vec r_1)
 \Psi_d(\vec r_4)\left[\phi_{2s}(\vec r_2)\Psi_{mol}(\vec r_3,\vec r_\mu)+
 \phi_{2s}(\frac{\vec r_2}{3}-\vec r_3)\Psi_{mol}(-\frac{5\vec r_2}{6}-
 \frac{\vec r_3}{2},\vec r_\mu)\right]\xi_{(12)(34)}}\right.
\nonumber\\
 &&-\Psi_d(\frac{\vec r_1}{2}+\frac{2\vec r_2}{3}+\vec r_3+
  \frac{\vec r_4}{2}) \Psi_d(\frac{\vec r_1}{2}-
   \frac{2\vec r_2}{3}-\vec r_3+\frac{\vec r_4}{2})
\nonumber\\
  &&\times\left[\phi_{2s}(-\frac{\vec r_1}{4}+\frac{2\vec r_2}{3}
    -\frac{\vec r_3}{2}+ \frac{\vec r_4}{4})\right.
     \Psi_{mol}(-\frac{\vec r_1}{3}-\frac{4\vec r_2}{9}+
   \frac{\vec r_3}{3}+\frac{\vec r_4}{3},\vec r_\mu)
\nonumber\\
   &&+ \left.
    \phi_{2s}(\frac{\vec r_1}{4}+\frac{2\vec r_2}{3}-\frac{\vec r_3}{2}-
    \frac{\vec r_4}{4})\Psi_{mol}(\frac{\vec r_1}{3}-\frac{4\vec r_2}{9}
    +\frac{\vec r_3}{3}-\frac{\vec r_4}{3},\vec r_\mu)\right]\xi_{(14)(32)}
\nonumber\\
 &&-\Psi_d(\frac{\vec r_1}{2}-\frac{2\vec r_2}{3}-\vec r_3-
              \frac{\vec r_4}{2})
   \Psi_d(-\frac{\vec r_1}{2}-\frac{2\vec r_2}{3}-\vec r_3-
             \frac{\vec r_4}{2})
\nonumber\\
  &&\times\left[\phi_{2s}(-\frac{\vec r_1}{4}+\frac{2\vec r_2}{3}-
      \frac{\vec r_3}{2}-\frac{\vec r_4}{4})
    \Psi_{mol}(-\frac{\vec r_1}{3}-\frac{4\vec r_2}{9}+
    \frac{\vec r_3}{3}-\frac{\vec r_4}{3},\vec r_\mu)\right.
\nonumber\\
 &&+\left.{\left.\phi_{2s}(\frac{\vec r_1}{4}+\frac{2\vec r_2}{3}
     -\frac{\vec r_3}{2}+ \frac{\vec r_4}{4})
    \Psi_{mol}(\frac{\vec r_1}{3}-\frac{4\vec r_2}{9}+\frac{\vec r_3}{3}+
     \frac{\vec r_4}{3},\vec r_\mu)\right]}\xi_{(13)(24)}\right\}\,.
\nonumber
\end{eqnarray}
The normalization constant ${\cal N}_1$ is obtained numerically.
%%%%%%%%%%%%%%%%%%%%%%%%%%%%%%%%%%%%%%%%%%%%%%%%%%%%%%%%%%%%%%%%%%%%%%
%%%%%%%%  POTENTIALS AND MATRIX ELEMENTS  %%%%%%%%%%%%%%%%%%%%%%%%%%%
%%%%%%%%%%%%%%%%%%%%%%%%%%%%%%%%%%%%%%%%%%%%%%%%%%%%%%%%%%%%%%%%%%%%%%
\section{Potentials and matrix elements}

For the description of the strong interactions among the particles of
the five-body  $(NNNN\alpha)$  system we need the  nucleon-nucleon,
$v_{mn}^{NN}$, and nucleon-$\alpha, v_{m\alpha}^{N\alpha}$, potentials
$(m,n=1,2,3,4)$.\\

Since $\Psi_1$ and $\Psi_2$  are constructed from  $S$-wave components only,
the total angular momentum $J=2$ of the $(NNNN\alpha)$ system
can be formed  from the spin-momenta of the four nucleons. This  is
possible when the spins are aligned, i.e. when the total spin
of any $NN$-pair equals 1. Thus, as a nucleon-nucleon potential, we employ
the Malfliet-Tjon triplet potential \cite{mt},
\begin{equation}
     v^{NN}(r)=V_1\frac{e^{-\alpha_1r}}{r}-V_2\frac{e^{-\alpha_2r}}{r},
\end{equation}
with
    $V_1=1438.72\,{\rm MeV\,fm}$,  $V_2=626.885\,{\rm MeV\,fm}$,
   $\alpha_1=3.11\,{\rm fm}^{-1}$, and $ \alpha_2=1.55\,{\rm fm}^{-1}$.\\

For the $N\alpha$ potential, we choose the one proposed in Ref. \cite{nalpha}
\begin{equation}
v^{N\alpha}(r)=-V_0\exp(-wr^2)
\end{equation}
where $V_0=55.774$ MeV and $w=0.292$ fm$^{-2}$.\\

For the matrix elements we have
\begin{equation}
   V_{ij}^s\equiv\langle\Psi_i|v_{13}^{NN}+v_{14}^{NN}+v_{23}^{NN}+
   v_{24}^{NN} +v_{3\alpha}^{N\alpha}+v_{4\alpha}^{N\alpha}|\Psi_j\rangle.
\end{equation}
Due to the symmetry properties of $\Psi_1$ and $\Psi_2$ with respect to
nucleon permutations, we may write
\begin{equation}
\label{vsij}
V_{ij}^s=4\langle\Psi_i|v_{14}^{NN}|\Psi_j\rangle+2\langle\Psi_i|
v_{4\alpha}^{N \alpha}|\Psi_j\rangle.
 \end{equation}
 After antisymmetrization, the wave functions $\Psi_1$ and $\Psi_2$
 acquire components with nonzero orbital angular momenta. However, in
 the integrals (\ref{vsij}) the $S$-wave potential operators which
 implicitly include the  projection operators ${\cal P}(\ell_{14}=0)$ and
 ${\cal P}(\ell_{4\alpha}=0)$ which retain only $S$-waves  along the
vectors
 \begin{eqnarray}
          \vec x&=&\frac{\vec r_1}{2}+\frac{2\vec r_2}{3}+\vec r_3
              +\frac{\vec r_4}{2},
 \nonumber\\
\phantom{-}\label{auxx}\\
          \vec y&=&-\frac{\vec r_2}{3}+\vec r_3+\frac{\vec r_4}{2},
 \nonumber
 \end{eqnarray}
 towards the first nucleon and the $\alpha$-particle from the position of
 the fourth nucleon respectively (see Fig. 3 ). In order to perform  this
$S$-wave  projection we make the additional integration
 \begin{eqnarray}
        {\cal P}(\ell_{14}=0)\Psi_i=\frac{1}{4\pi}\int d\hat{\bf x}
           \Psi_i\equiv\Phi_{ix},
 \nonumber\\
        {\cal P}(\ell_{4\alpha}=0)\Psi_i=\frac{1}{4\pi}\int d\hat{\bf y}
        \Psi_i\equiv\Phi_{iy}.
\nonumber
\end{eqnarray}
Then the matrix elements are calculated as follows
\begin{eqnarray}
       \langle\Psi_i|v_{14}^{NN}|\Psi_j\rangle&=&(4\pi)^2\int
       \limits_0^\infty dr_1dx(r_1x)^2\int d\vec r_2d\vec r_4 d\vec r_{\mu}
       \Phi_{ix}^*v_{14}^{NN}\Phi_{jx},
\label{V14}\\
      \langle\Psi_i|v_{4\alpha}^{N\alpha}|\Psi_j\rangle&=&(4\pi)^2\int
      \limits_0^\infty dr_1 dy(r_1y)^2\int d\vec r_2d\vec r_4 d\vec r_{\mu}
      \Phi_{iy}^*v_{4\alpha}^{N\alpha}\Phi_{jy}.\label{V4A}
\end{eqnarray}
For the overlapping integral $I$ we have
\begin{equation}
\label{OV}
    I=4\pi\int\limits_0^\infty dr_1r_1^2\int d\vec r_2d\vec r_3d\vec r_4
    d\vec r_{\mu} \Psi_1^*(\vec r_1,\vec r_2,\vec r_3,\vec r_4,\vec r_\mu)
    \Psi_{^8\!Be^*}(\vec r_1,\vec r_4,\frac{2\vec r_2}{3}+\vec r_3,\frac{2
    \vec r_2}{3}-\frac{\vec r_3}{2})\phi_\mu(\vec r_\mu)\,,
\end{equation}
where $\vec r_1$ is taken along  the  $z$-axis of the reference frame for the
vectors $\vec r_2,\,\vec r_3,$ and $\vec r_4$. The dot products among
these vectors, appearing in the integrand have the general structure
$$
\vec r_m\cdot\vec r_n=r_mr_n\lbrack\sin\theta_m\sin\theta_n\cos(\varphi_m-
\varphi_n)
+\cos\theta_m\cos\theta_n\rbrack,
$$
where $m,n=1,2,3,4$ and $\theta_m,\varphi_m$ are the spherical angles of
$\vec r_m$ in that frame. The combination $\Psi_1^*\Psi_{^8\!Be^*}$ contains
dot products among the spin-isospin functions
$$
     \xi_1\equiv\xi_{(12)(34)},\quad\xi_2\equiv\xi_{(14)(32)},\quad{\rm and}
     \quad\xi_3\equiv\xi_{(13)(24)}.
$$
It is easily seen that $\langle\xi_k|\xi_l\rangle=1/2$ for all
$k$ and $l$.\\

The above multidimensional integrals (\ref{V14}), (\ref{V4A}),
 and (\ref{OV}) involve an integration over the muon variable
$\vec r_\mu$ (see Fig. 3). In order to reduce their dimensions, we exploit
 the fact that nuclear wave functions and potentials are localized within
 a small  volume of the size  which is not lager than few fm, while
 the molecular wave function $\Psi_{mol}(\vec R,\vec r_{\mu})$ is practically
constant when $R$ is within a region of $\sim 100$ fm from the point $R=0$.\\

Thus, in these integrals we can replace $\Psi_{mol}(\vec R,\vec r_\mu)$ by
$\Psi_{mol}(0,\vec r_\mu)$. On the other hand, $\Psi_{mol}(0,\vec r_\mu)$
is the so-called united atom limit of the three-body molecular wave
function and describes the motion of the muon in the joint Coulomb field of
the two nuclei with zero separation between them. Hence, in this limit we
have
\begin{equation}
\Psi_{mol}(\vec R,\vec r_\mu)\mathop{\longrightarrow}\limits_{R\to0}
\Psi_m(0)\phi_\mu(\vec r_\mu)\,,
\label{fact}
\end{equation}
where $\phi_\mu(\vec r_\mu)$ is the atomic wave function, and $\Psi_m(0)$
is the limit $(R\to0)$ of the wave function describing the nucleus-nucleus
relative motion inside the molecule ( the calculated value of $\Psi_m(0)$
was given in Sec. III ).\\

Therefore, by substituting the factorized expression (\ref{fact}) into
the above integrals, the integration over the muon variable $\vec r_\mu$ is
eliminated. The most formidable of the remaining integrals is one for the
overlapping $I$ which still has 10 dimensions.

Finally, we have to find the matrix element $H_{22}^\mu$ of the
Hamiltonian $H^\mu(\vec r_1,\vec r_2,\vec r_3,\vec r_4)$ describing the muon
motion in the Coulomb field generated by  the nucleons. This
Hamiltonian depends on the nucleon variables parametrically. Thus we have
$$
H_{22}^\mu=\langle\Psi_{^8\!Be^*}\phi_\mu|H^\mu
(\vec r_1,\vec r_2,\vec r_3,\vec r_4)|\Psi_{^8\!Be^*}\phi_\mu\rangle\,,
$$
and again, exploiting the smallness of the nuclear size, we use the reliable
approximation
\begin{eqnarray}
H_{22}^\mu & \approx & \langle\Psi_{^8\!Be^*}\phi_\mu|H^\mu(0,0,0,0)
|\Psi_{^8\!Be^*}\phi_\mu\rangle
\nonumber\\
&&=\langle \phi_\mu|H^\mu(0,0,0,0)|\phi_\mu\rangle\,.
\nonumber
\end{eqnarray}
The last matrix element is just the eigenenergy of the muonic atom
corresponding to the eigenstate $|\phi_\mu\rangle$ with the quantum numbers
$(n\ell m)$. This eigenenergy is the one of a hydrogen-like atom and is
expressed via the grand orbital quantum number $n$ as
$-(4/n)^2Ry$ where $Ry=m_\mu e^4/2\hbar^2=2813.25\,{\rm eV}$
is the Rydberg constant for muon.\\

 A question then arizes: which of the atomic states should be used here?
 The answer can be found with the help of the Born-Oppenheimer model. Indeed,
in this model for the hydrogen-lithium system \cite{pon}, only the so-called
$3d\sigma$-term provides the shallow well shown in  Fig. 1.
The solutions of the two-center Coulomb problem are usually classified by
 the quantum numbers $(n\ell m)$ of the united atom, $(R\to0)$, which is
formed when the muon moves in the Coulomb field of the total charge of the
nuclei. Thus, the symbol $3d\sigma$ means that, when the nuclei approach
each other, the muon is in the atomic state with  grand orbital quantum
 number $n=3$, angular momentum $\ell=2$, and $m=0$.
Moreover, when $R\to\infty$ the $3d\sigma$-term correspond to free motion of
lithium nucleus and $d\mu$-atom in its ground state. Hence, as the zero
point of the energy scale we should choose the ground state energy, $(-Ry)$,
of $d\mu$-atom. Therefore
$$
     H_{22}^\mu\approx -\left(\frac43\right)^2Ry+Ry=-\frac79 Ry.
$$
We epmhasize that the Born-Oppenheimer model was used here only to
find the  quantum numbers of the atoms which are formed in the limits
$R\to0$ and $R\to\infty$.
%%%%%%%%%%%%%%%%%%%%%%%%%%%%%%%%%%%%%%%%%%%%%%%%%%%%%%%%%%%%%%%%%%%%%%%%%
%%%%%%%%%%%%%%%%%%  RESULTS AND DISCUSSION    %%%%%%%%%%%%%%%%%%%%%%%%%%%
%%%%%%%%%%%%%%%%%%%%%%%%%%%%%%%%%%%%%%%%%%%%%%%%%%%%%%%%%%%%%%%%%%%%%%%%%
\section{Results and discussion}
%%%%%%%%%%%%%%%%%%%%%%%%%%%%%%%%%
The numerical evaluation of the multidimensional integrals for the potential
matrix elements and the overlapping integral, were performed with the help of
the Haar-function expansion method \cite{haar}. The results thus obtained
are
\begin{eqnarray}
\nonumber
      V_{11}^s&=&-0.14524\;10^{-2}\,{\rm eV},\\
\nonumber
     V_{12}^s&=&V_{21}^s=-6.2484\,{\rm eV},\\
\nonumber
     H_{22}^\mu&=&-5001.3\,{\rm eV},\\
\nonumber
     I&=&0.42501\;10^{-6}.
\end{eqnarray}
The use of these matrix elements in the formulae of Sec. II along with
$$
   E_1=22.2798\,{\rm MeV}\quad{\rm and}\quad E_2=22.2\,
    {\rm MeV}
$$
gives for the molecular level displacement $\Delta_1$, transition
probability $P$, and  reaction rate $\lambda$ the following results:
\begin{eqnarray}
     \Delta_1&=&-0.963\;10^{-3}\;{\rm eV},
\nonumber\\
     P&=&0.606\;10^{-8},\label{rez}\\
   \lambda&=&0.298\;10^8\;{\rm sec}^{-1}.
\nonumber
\end{eqnarray}
At this point it is important to note that the compound nuclear state
$^8\!Be^*(2^{+},0)$ is a resonant one with  energy, $(22.2-i0.8)$ MeV,
having a rather large width $\Gamma=1.6$ MeV. This width is much greater
 than the assumed difference,
$$
         \delta=E_1-E_2=79800\;{\rm eV},
$$
between the molecular and compound nucleus levels. We further note that
a resonance can be excited within a wide energy interval around
 the central resonant energy \cite{kuk}, and the relative probability
of its excitation is defined by the Breit-Wigner factor
$$
          W(E)=\frac{\Gamma/2}{(E-E_2)^2+\Gamma^2/4}.
$$
Therefore, in order to obtain the correct values of $\Delta_1,P$,
and $\lambda$, one should calculate them with
different values of $\delta=E_1-E_2$ and take the average value
with $W(E)$ as the weight factor.\\

The dependence of $\Delta_1,P$, and $\lambda$ on the energy level
difference $\delta$ is shown in Table \ref{t1}. It is seen that the
reaction rate $\lambda$ significantly increases when $\delta\to0$,
and attains a very high value $0.246\;10^{16}\;{\rm sec}^{-1}$
when the levels $E_1$ and $E_2$ coincide. To include these contributions
we note that  the nuclear transition (\ref{reac}) can only take place if
$E_1\ge E_2$. Further, since $\delta$ cannot be greater than $E_1$,
 the interval of $\delta$ for the averaging is $[0,E_1]$. Thus,
the average value of the reaction rate is  obtained via
\begin{equation}
     \langle\lambda\rangle=\frac{\int\limits_0^{E_1}d\delta\lambda(\delta)
     W(E_1-\delta)}{\int\limits_0^{E_1}d\delta W(E_1-\delta)}.
\label{averl}
\end{equation}
The averages $\langle\Delta_1\rangle$ and $\langle P \rangle$ ,
 can be similarly defined. The  results which take into account the
spreading of the final resonant state are
\begin{eqnarray}
        \langle\Delta_1\rangle&=&-0.116\;10^{-2}\;{\rm eV},
\nonumber\\
        \langle P\rangle&=&0.373\;10^{-6},\nonumber\\
       \langle\lambda\rangle&=&0.183\;10^{10}\;{\rm sec}^{-1}.
\nonumber
\end{eqnarray}
Our conclusions can be summarized as follows. We employed the LCAO-motivated
approach to calculate the transition probability and reaction rate for the
nuclear fusion inside the muonic $d-\mu-{^6\!Li}$ molecule by using the fact
that $^8\!Be$ has an excited state near the $d\,{^6\!L}i$ threshold energy.
The wave functions of the nuclear subsystems in these calculations were
constructed from antisymmetrized products of the harmonic oscillator
functions  in a three cluster ($dd\alpha$) approximation to the
five-body model ($NNNN\alpha$) while the three-body molecular problem was
treated in the framework of the complex coordinate rotation method.\\

Within this model, the reaction rate strongly depends on the gap
between the initial and final state energies. Since, however, the final
state  in our case is a resonance, we consider it reasonable to average the
results over the permissible (by the reaction)  energies covered by the
resonance. The average results turned out to be two orders of magnitude
higher as compared to those obtained using the center of the resonance.
This makes the reaction (1) quite  attractive for further
theoretical as well as experimental investigations.\\

%%%%%%%%%%%%%%%%%%%%%%%%%%%%%%%%%%%%%%%%%%%%%%%%%%%%%%%%%%%%%%%%%%%
%%%%%%%%%%%%%%%%%   ACKNOWLEDGEMENTS   %%%%%%%%%%%%%%%%%%%%%%%%%%%
%%%%%%%%%%%%%%%%%%%%%%%%%%%%%%%%%%%%%%%%%%%%%%%%%%%%%%%%%%%%%%%%%%%
\begin{center}{\large Acknowledgements}\end{center}
Financial support from the  University of South
Africa  and the Joint Institute  for Nuclear Research, Dubna,
is greately acknowledged. This work  was also partly supported by
the Grants RFB300 and NSCT000 of the International Science Foundation and
Russian Government, and by the NATO Collaborative Research Grant N930102.
%%%%%%%%%%%%%%%%%%%%%%%%%%%%%%%%%%%%%%%%%%%%%%%%%%%%%%
\newpage
%%%%%%%%%%%%%%%%%%%%%%%%%%%%%%%%%%%%%%%%%%%%%%%
%%%%%%%%%%%%%%   REFERENCES  %%%%%%%%%%%%%%%%%%
%%%%%%%%%%%%%%%%%%%%%%%%%%%%%%%%%%%%%%%%%%%%%%%

\newpage
\begin{table*}
\begin{tabular}{c@{\hspace{5mm}}c@{\hspace{5mm}}c@{\hspace{5mm}}c}
%\hline\hline
$\delta\;{\rm (eV)}$ & $\Delta_1\;{\rm (eV)}$ & $P$ &
$\lambda\;({\rm sec})^{-1}$\\
\hline
0 & -6.25 & 0.499 & $0.246\;10^{16}$\\
1 & 5.77 & 0.460 & $0.226\;10^{16}$\\
10 & 3.00 & 0.188 & $0.921\;10^{15}$\\
$10^2$ & 0.387 & $0.386\;10^{-2}$ & $0.190\;10^{14}$\\
$10^3$ & $0.376\;10^{-1}$ & $0.390\;10^{-4}$ & $0.192\;10^{12}$\\
$10^4$ & $0.245\;10^{-2}$ & $0.390\;10^{-6}$ & $0.191\;10^{10}$\\
${\bf 79800}$ & ${\bf-0.963\;10^{-3}}$ & ${\bf 0.606\;10^{-8}}$
& ${\bf 0.298\;10^{8}}$\\
$10^5$ & $-0.106\;10^{-2}$ & $0.385\;10^{-8}$ & $0.189\;10^{8}$\\
$10^6$ & $-0.141\;10^{-2}$ & $0.339\;10^{-10}$ & $0.167\;10^{6}$\\
%\hline\hline
\end{tabular}

\caption{ The dependence of the molecular level displacement $\Delta_1$,
transition probability $P$, and reaction rate $\lambda$ on the energy
difference $\delta$ between the molecular $E_1$ and compound-nucleus levels.}
\label{t1}
\end{table*}
%%%%%%%%%%%%%%%%%%%%%%%%%%%%%%%%%%%%%%%%%%%%%%%%%%%%%%%%%%%%%%%%%%%%%%%%
%%%%%%%%%%%%%%%%%%%%%  FIGURES %%%%%%%%%%%%%%%%%%%%%%%%%%%%%%%%%%%%%%%%%

\def\emline#1#2#3#4#5#6{%
      \put(#1,#2){\special{em:moveto}}
       \put(#4,#5){\special{em:lineto}}}
\def\newpic#1{}
%%%%%%%%%%%%%%%%%%%%%%%%%%%%%%%%%%%%%%%%%%%%%%%%%%%%%%%%%%
%%%%%%%%%%%%%%%%% POTENTIAL  {fig1}%%%%%%%%%%%%%%%%%%%%%%%%%%%%%%%%%%

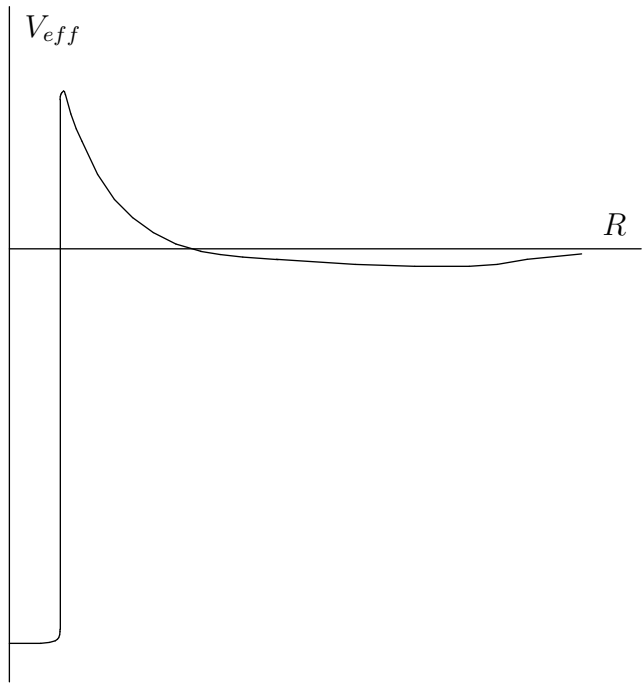
\begin{figure}
\label{fig1}
%\centering
%
\unitlength=0.70mm
\special{em:linewidth .5pt}
\linethickness{.5pt}
\begin{picture}(145.00,151.00)
\emline{25.00}{105.00}{1}{145.00}{105.00}{2}
\emline{31.00}{30.00}{3}{25.00}{30.00}{4}
\emline{75.67}{103.00}{5}{90.67}{102.00}{6}
\emline{90.67}{102.00}{7}{102.00}{101.67}{8}
\emline{102.00}{101.67}{9}{112.33}{101.67}{10}
\emline{112.33}{101.67}{11}{117.67}{102.00}{12}
\emline{117.67}{102.00}{13}{123.33}{103.00}{14}
\emline{123.33}{103.00}{15}{133.67}{104.00}{16}
\put(137.67,107.67){\makebox(0,0)[lb]{$R$}}
\emline{35.67}{134.30}{17}{36.67}{130.64}{18}
\emline{35.00}{134.64}{19}{35.28}{134.96}{20}
\emline{35.28}{134.96}{21}{35.49}{134.84}{22}
\emline{35.49}{134.84}{23}{35.65}{134.27}{24}
\emline{36.66}{130.62}{25}{37.71}{127.79}{26}
\emline{37.71}{127.79}{27}{38.81}{125.40}{28}
\emline{38.78}{125.43}{29}{41.77}{119.15}{30}
\emline{41.77}{119.15}{31}{44.96}{114.37}{32}
\emline{44.96}{114.37}{33}{48.45}{110.88}{34}
\emline{48.45}{110.88}{35}{52.34}{108.08}{36}
\emline{52.34}{108.08}{37}{56.63}{105.89}{38}
\emline{56.63}{105.89}{39}{61.41}{104.49}{40}
\emline{61.30}{104.52}{41}{65.07}{103.89}{42}
\emline{65.07}{103.89}{43}{69.32}{103.40}{44}
\emline{69.32}{103.40}{45}{76.08}{102.91}{46}
\emline{25.00}{151.00}{47}{25.00}{22.67}{48}
\put(28.00,145.33){\makebox(0,0)[lb]{$V_{eff}$}}
\emline{34.67}{133.33}{49}{34.67}{32.67}{50}
\emline{34.66}{133.32}{51}{34.72}{133.93}{52}
\emline{34.72}{133.93}{53}{34.83}{134.35}{54}
\emline{34.83}{134.35}{55}{35.02}{134.66}{56}
\emline{30.96}{29.99}{57}{32.64}{30.14}{58}
\emline{32.64}{30.14}{59}{33.70}{30.48}{60}
\emline{33.70}{30.48}{61}{34.17}{30.75}{62}
\emline{34.17}{30.75}{63}{34.42}{31.07}{64}
\emline{34.42}{31.07}{65}{34.60}{31.60}{66}
\emline{34.60}{31.60}{67}{34.64}{32.20}{68}
\emline{34.64}{32.20}{69}{34.66}{32.67}{70}

\caption{ Typical structure of the effective two-body interaction
potential acting between the nuclei confined within a molecule.}
\end{picture}
\end{figure}
%%%%%%%%%%%%%%%%%%%%%%%%%%%%%%%%%%%%%%%%%%%%%%%%%%%%%%%%%%%%
%%%%%%%%%%%%%%%%%%%%%% TRAJECTORIES  {fig2} %%%%%%%%%%%%%%%%%
\begin{figure}
\label{fig2}
\unitlength=1.10mm
\special{em:linewidth .5pt}
\linethickness{.5pt}
\begin{picture}(120.00,145.67)
\emline{20.00}{140.00}{1}{120.00}{140.00}{2}
\emline{120.00}{140.00}{3}{120.00}{80.00}{4}
\emline{120.00}{80.00}{5}{20.00}{80.00}{6}
\emline{20.00}{80.00}{7}{20.00}{140.00}{8}
\emline{20.00}{120.00}{9}{17.00}{120.00}{10}
\emline{17.00}{100.00}{11}{20.00}{100.00}{12}
\emline{40.00}{80.00}{13}{40.00}{77.00}{14}
\emline{60.00}{80.00}{15}{60.00}{77.00}{16}
\emline{80.00}{80.00}{17}{80.00}{77.00}{18}
\emline{100.00}{80.00}{19}{100.00}{77.00}{20}
\emline{100.00}{140.00}{21}{100.00}{137.00}{22}
\emline{80.00}{140.00}{23}{80.00}{137.00}{24}
\emline{60.00}{140.00}{25}{60.00}{137.00}{26}
\emline{40.00}{140.00}{27}{40.00}{137.00}{28}
\put(120.00,120.00){\rule{-3.00\unitlength}{0.00\unitlength}}
\emline{120.00}{120.00}{29}{117.00}{120.00}{30}
\emline{120.00}{100.00}{31}{117.00}{100.00}{32}
\emline{20.00}{140.00}{33}{17.00}{140.00}{34}
\emline{20.00}{80.00}{35}{17.00}{80.00}{36}
\put(15.00,140.00){\makebox(0,0)[rc]{$0.00$}}
\put(15.00,120.00){\makebox(0,0)[rc]{$-0.40$}}
\put(15.00,100.00){\makebox(0,0)[rc]{$-0.80$}}
\put(15.00,80.00){\makebox(0,0)[rc]{$-1.20$}}
\put(20.67,145.67){\makebox(0,0)[cb]{$I\!m\,E\,(10^{-2}{\rm eV})$}}
\put(40.00,75.00){\makebox(0,0)[ct]{$-20.313$}}
\put(60.00,75.00){\makebox(0,0)[ct]{$-20.311$}}
\put(80.00,75.00){\makebox(0,0)[ct]{$-20.309$}}
\put(100.00,75.00){\makebox(0,0)[ct]{$-20.307$}}
\put(105.00,67.00){\makebox(0,0)[lc]{$R\!e\,E\,({\rm eV})$}}
\put(34.00,140.00){\circle{2.00}}
\put(56.00,114.00){\circle{2.11}}
\put(76.00,108.00){\circle{2.00}}
\put(82.00,107.00){\circle{2.11}}
\put(84.67,107.00){\circle{2.11}}
\put(81.00,89.00){\circle{2.00}}
\put(83.00,98.00){\circle{1.89}}
\put(84.00,102.67){\circle{2.11}}
\put(84.67,105.67){\circle{2.00}}
\emline{80.97}{88.93}{37}{83.07}{98.03}{38}
\emline{83.07}{98.03}{39}{84.12}{102.73}{40}
\emline{84.12}{102.73}{41}{84.78}{105.71}{42}
\emline{84.78}{105.71}{43}{84.67}{107.03}{44}
\emline{84.67}{107.03}{45}{82.08}{106.92}{46}
\emline{82.08}{106.92}{47}{75.78}{108.14}{48}
\emline{76.00}{108.00}{49}{56.00}{114.00}{50}
\emline{56.00}{114.00}{51}{34.00}{140.00}{52}
\put(43.33,132.33){\makebox(0,0)[lc]{$\alpha=1.00$}}
\put(86.00,140.00){\circle*{2.00}}
\put(85.67,121.00){\circle*{1.89}}
\put(84.33,113.00){\circle*{2.00}}
\put(84.67,109.33){\circle*{2.00}}
\put(85.33,106.33){\circle*{2.00}}
\put(104.00,92.00){\circle*{2.11}}
\put(94.00,100.00){\circle*{2.00}}
\put(89.67,103.33){\circle*{2.00}}
\put(87.33,105.00){\circle*{2.00}}
\emline{103.97}{91.95}{53}{94.00}{99.95}{54}
\emline{94.00}{99.95}{55}{89.70}{103.31}{56}
\emline{89.70}{103.31}{57}{87.38}{105.17}{58}
\emline{87.38}{105.17}{59}{85.30}{106.33}{60}
\emline{85.30}{106.33}{61}{84.83}{109.69}{62}
\emline{84.83}{109.69}{63}{84.37}{113.17}{64}
\emline{84.37}{113.17}{65}{85.76}{121.18}{66}
\emline{85.76}{121.18}{67}{85.99}{139.97}{68}
\put(39.00,80.67){\rule{1.67\unitlength}{1.67\unitlength}}
\put(59.33,93.67){\rule{1.67\unitlength}{1.33\unitlength}}
\put(69.33,99.67){\rule{1.67\unitlength}{1.33\unitlength}}
\put(75.67,103.33){\rule{1.67\unitlength}{1.33\unitlength}}
\put(80.67,105.00){\rule{1.67\unitlength}{1.33\unitlength}}
\put(83.67,106.00){\rule{1.00\unitlength}{0.67\unitlength}}
\put(81.00,102.33){\rule{1.67\unitlength}{1.33\unitlength}}
\put(76.67,99.33){\rule{1.67\unitlength}{1.33\unitlength}}
\put(60.00,88.67){\rule{1.67\unitlength}{1.33\unitlength}}
\emline{39.75}{81.59}{69}{60.25}{94.40}{70}
\emline{60.25}{94.40}{71}{70.15}{100.46}{72}
\emline{70.15}{100.46}{73}{76.56}{104.07}{74}
\emline{76.56}{104.07}{75}{81.56}{105.58}{76}
\emline{81.56}{105.58}{77}{84.47}{106.74}{78}
\emline{84.47}{106.74}{79}{81.91}{102.90}{80}
\emline{81.91}{102.90}{81}{77.49}{99.87}{82}
\emline{77.49}{99.87}{83}{60.60}{89.28}{84}
\put(88.67,130.00){\makebox(0,0)[lc]{$\alpha=0.98$}}
\put(44.33,87.33){\makebox(0,0)[rc]{$\alpha=1.02$}}
\caption{
Complex eigenvalue trajectories corresponding to variations of the
rotation angle $\theta_i$ and for different values of dilatation parameter
$\alpha=e^{\theta_r}$.}
\end{picture}
\end{figure}
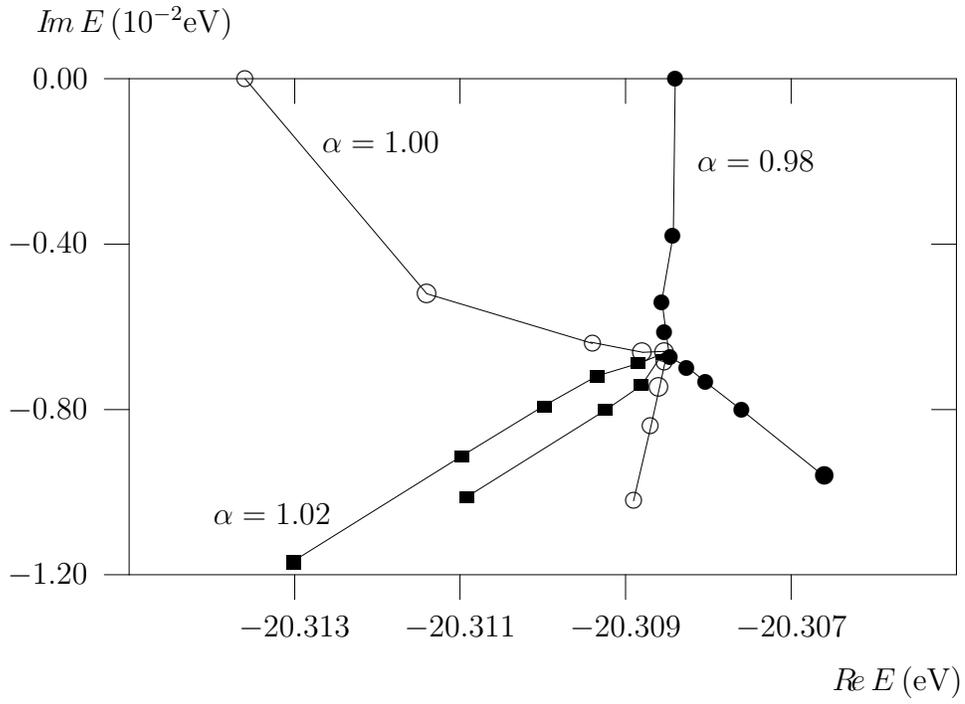
%%%%%%%%%%%%%%%%%%%%%%%%%%%%%%%%%%%%%%%%%%%%%%%%%%%%%%%%%%%%
%%%%%%%%%%%%%%%%%%%% JACOBI VECTORS {fig3} %%%%%%%%%%%%%%%%%
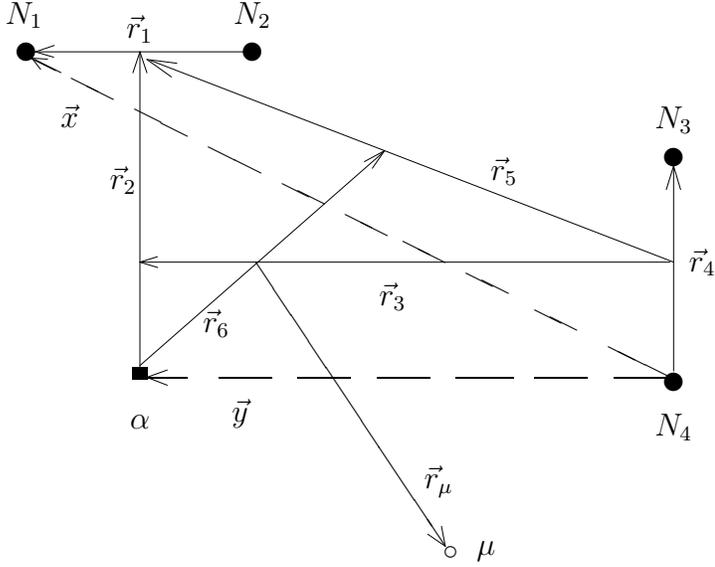
\begin{figure}
\label{fig3}
\unitlength=0.70mm
\special{em:linewidth 0.5pt}
\linethickness{0.5pt}
\begin{picture}(139.33,132.00)
\emline{35.00}{125.00}{1}{35.00}{65.00}{2}
\emline{35.00}{85.00}{3}{135.67}{85.00}{4}
\emline{136.33}{105.00}{5}{136.33}{64.33}{6}
\emline{15.00}{125.00}{7}{55.00}{125.00}{8}
\put(56.33,125.00){\circle*{3.40}}
\put(136.33,105.00){\circle*{3.40}}
\put(136.33,62.33){\circle*{3.40}}
\put(13.33,125.00){\circle*{3.40}}
\put(33.67,62.67){\rule{2.67\unitlength}{2.33\unitlength}}
\emline{36.33}{121.00}{9}{35.00}{125.00}{10}
\emline{35.00}{125.00}{11}{33.67}{121.00}{12}
\emline{37.00}{123.33}{13}{136.33}{85.00}{14}
\emline{81.33}{106.00}{15}{35.00}{65.33}{16}
\put(13.00,132.00){\makebox(0,0)[cc]{$N_1$}}
\put(56.33,132.00){\makebox(0,0)[cc]{$N_2$}}
\put(136.33,112.00){\makebox(0,0)[cc]{$N_3$}}
\put(136.33,53.67){\makebox(0,0)[cc]{$N_4$}}
\put(35.00,54.67){\makebox(0,0)[cc]{$\alpha$}}
\put(35.00,129.33){\makebox(0,0)[cc]{$\vec r_1$}}
\put(29.33,100.67){\makebox(0,0)[lc]{$\vec r_2$}}
\put(83.00,81.00){\makebox(0,0)[ct]{$\vec r_3$}}
\put(139.33,85.00){\makebox(0,0)[lc]{$\vec r_4$}}
\put(101.67,100.67){\makebox(0,0)[lb]{$\vec r_5$}}
\put(52.00,73.33){\makebox(0,0)[rc]{$\vec r_6$}}
\emline{77.63}{104.14}{17}{81.43}{106.15}{18}
\emline{81.43}{106.15}{19}{79.17}{102.83}{20}
\emline{38.35}{86.06}{21}{34.99}{85.00}{22}
\emline{34.99}{85.00}{23}{38.38}{84.00}{24}
\emline{134.95}{98.81}{25}{136.37}{103.25}{26}
\emline{136.37}{103.25}{27}{137.71}{98.78}{28}
\emline{40.98}{120.19}{29}{36.34}{123.52}{30}
\emline{36.34}{123.52}{31}{42.02}{122.79}{32}
\put(136.00,63.00){\line(-2,1){12.02}}
\put(135.00,63.00){\line(-1,0){10.00}}
\put(120.00,63.00){\line(-1,0){10.00}}
\put(106.00,63.00){\line(-1,0){11.00}}
\put(90.00,63.00){\line(-1,0){10.00}}
\put(75.00,63.00){\line(-1,0){10.00}}
\put(60.00,63.00){\line(-1,0){10.00}}
\put(54.00,59.00){\makebox(0,0)[ct]{$\vec y$}}
\put(118.00,72.00){\line(-2,1){8.00}}
\put(105.00,79.00){\line(-2,1){8.00}}
\put(93.00,85.00){\line(-2,1){8.00}}
\put(82.00,90.00){\line(-2,1){8.00}}
\put(70.00,96.00){\line(-2,1){8.00}}
\put(58.00,102.00){\line(-2,1){10.00}}
\put(44.00,109.00){\line(-2,1){10.00}}
\put(30.00,116.00){\line(-2,1){8.00}}
\put(20.00,115.00){\makebox(0,0)[lt]{$\vec x$}}
\put(94.00,30.00){\circle{2.00}}
\put(89.00,42.00){\makebox(0,0)[lb]{$\vec r_\mu$}}
\put(99.00,30.00){\makebox(0,0)[lc]{$\mu$}}
\put(57.00,85.00){\line(2,-3){36.00}}
\put(44.00,63.00){\line(-1,0){8.02}}
\put(20.00,121.00){\line(-2,1){6.00}}
\emline{92.26}{34.80}{33}{93.02}{31.03}{34}
\emline{93.02}{31.03}{35}{89.67}{33.29}{36}
\emline{40.17}{64.50}{37}{36.35}{63.10}{38}
\emline{36.35}{63.10}{39}{40.17}{61.59}{40}
\emline{18.74}{126.17}{41}{15.16}{124.96}{42}
\emline{15.16}{124.96}{43}{18.60}{123.98}{44}
\emline{17.44}{123.58}{45}{14.49}{123.75}{46}
\emline{14.49}{123.75}{47}{16.10}{121.65}{48}

\caption{
Two alternative sets of Jacobi vectors
$\left\{\vec r_1,\vec r_2,\vec r_3,\vec r_4,\vec r_\mu\right\}$
and
$\left\{\vec r_1,\vec r_4,\vec r_5,\vec r_6,\vec r_\mu\right\}$
describing space configuration of the six-body system
$(NNNN\alpha\mu)$.
The $\vec x$ and $\vec y$ are auxiliary vectors defined by Eq. (43).
}
\end{picture}
\end{figure}

%%%%%%%%%%%%%%%%%%%%%%%%%%%%%%%%%%%%%%%%%%%%%%%%%%%%%%%%%%%%

\end{document}